\documentclass[12pt]{article}
\usepackage{amsmath,color}
\usepackage[dvips]{graphicx}
\input amssym

\def\red#1{{\color{red} #1}}

\begin{document}

\def\prg#1{\par\medskip\noindent{\bf #1}}  \def\ra{\rightarrow}
\newcounter{nbr}
\def\note#1{\bitem\vspace{-5pt}\addtocounter{nbr}{1}
            \item{} #1\vspace{-5pt}
            \eitem}
\def\lra{\leftrightarrow}              \def\Ra{\Rightarrow}
\def\nin{\noindent}                    \def\pd{\partial}
\def\dis{\displaystyle}                \def\Lra{{\Leftrightarrow}}
\def\grl{{GR$_\Lambda$}}               \def\vsm{\vspace{-8pt}}
\def\cs{{\scriptstyle\rm CS}}          \def\ads3{{\rm AdS$_3$}}
\def\Leff{\hbox{$\mit\L_{\hspace{.6pt}\rm eff}\,$}}
\def\bull{\raise.25ex\hbox{\vrule height.8ex width.8ex}}
\def\ric{{Ric}}                        \def\tmgl{\hbox{TMG$_\Lambda$}}
\def\Lie{{\cal L}\hspace{-.7em}\raise.25ex\hbox{--}\hspace{.2em}}
\def\sS{\hspace{2pt}S\hspace{-0.83em}\diagup}
\def\hd{{^\star}}                      \def\dis{\displaystyle}
\def\mb#1{\hbox{{\boldmath $#1$}}}     \def\kn#1{\hbox{KN$#1$}}
\def\ul#1{\underline{#1}}              \def\phb{\phantom{\Big|}}
\def\nb{~\marginpar{\bf\Large ?}}      \def\ph{\phantom{xxx}}

\def\hook{\hbox{\vrule height0pt width4pt depth0.3pt
\vrule height7pt width0.3pt depth0.3pt
\vrule height0pt width2pt depth0pt}\hspace{0.8pt}}
\def\inn{\hook}
\def\first{\rm (1ST)}  \def\second{\hspace{-1cm}\rm (2ND)}
\def\ppl{{pp${}_\Lambda$}}

\def\G{\Gamma}        \def\S{\Sigma}        \def\L{{\mit\Lambda}}
\def\D{\Delta}        \def\Th{\Theta}       \def\Ups{\Upsilon}
\def\a{\alpha}        \def\b{\beta}         \def\g{\gamma}
\def\d{\delta}        \def\m{\mu}           \def\n{\nu}
\def\th{\theta}       \def\k{\kappa}        \def\l{\lambda}
\def\vphi{\varphi}    \def\ve{\varepsilon}  \def\p{\pi}
\def\r{\rho}          \def\Om{\Omega}       \def\om{\omega}
\def\s{\sigma}        \def\t{\tau}          \def\eps{\epsilon}
\def\nab{\nabla}      \def\btz{{\rm BTZ}}   \def\heps{{\hat\eps}}

\def\bR{\bar{R}}      \def\bT{\bar{T}}     \def\hT{\hat{T}}
\def\tG{{\tilde G}}   \def\cF{{\cal F}}    \def\cA{{\cal A}}
\def\cL{{\cal L}}     \def\cM{{\cal M }}   \def\cE{{\cal E}}
\def\cH{{\cal H}}     \def\hcH{\hat{\cH}}  \def\cT{{\cal T}}
\def\hA{\hat{A}}      \def\hB{\hat{B}}     \def\hK{\hat{K}}
\def\cK{{\cal K}}     \def\hcK{\hat{\cK}}  \def\cT{{\cal T}}
\def\cO{{\cal O}}     \def\hcO{\hat{\cal O}} \def\cV{{\cal V}}
\def\tom{{\tilde\omega}}  \def\cE{{\cal E}} \def\bH{\bar{H}}
\def\cR{{\cal R}}    \def\hR{{\hat R}{}}   \def\hL{{\hat\L}}
\def\tb{{\tilde b}}  \def\tA{{\tilde A}}   \def\hom{{\hat\om}}
\def\tT{{\tilde T}}  \def\tR{{\tilde R}}   \def\tcL{{\tilde\cL}}
\def\he{{\hat e}}    \def\hom{{\hat\om}}   \def\hth{\hat\theta}
\def\hxi{\hat\xi}    \def\hg{\hat g}       \def\hb{{\hat b}}
\def\tH{{\tilde H}}  \def\tV{{\tilde V}}   \def\bA{\bar{A}}
\def\bV{\bar{V}}     \def\bxi{\bar{\xi}}
\def\knl{\text{KN}$(\l)$}   \def\mknl{\text{KN}\mb{(\l)}}
\def\bPhi{\bar\Phi}
\def\chm{\checkmark}                \def\chmr{\red{\chm}}
\vfuzz=2pt 
\def\nn{\nonumber}
\def\be{\begin{equation}}             \def\ee{\end{equation}}
\def\ba#1{\begin{array}{#1}}          \def\ea{\end{array}}
\def\bea{\begin{eqnarray} }           \def\eea{\end{eqnarray} }
\def\beann{\begin{eqnarray*} }        \def\eeann{\end{eqnarray*} }
\def\beal{\begin{eqalign}}            \def\eeal{\end{eqalign}}
\def\lab#1{\label{eq:#1}}             \def\eq#1{(\ref{eq:#1})}
\def\bsubeq{\begin{subequations}}     \def\esubeq{\end{subequations}}
\def\bitem{\begin{itemize}}           \def\eitem{\end{itemize}}
\renewcommand{\theequation}{\thesection.\arabic{equation}}

\title{Generalized pp waves in Poincar\'e gauge theory}

\author{M. Blagojevi\'c and B. Cvetkovi\'c\footnote{
        Email addresses: {\tt mb@ipb.ac.rs,
                          cbranislav@ipb.ac.rs}}\\
Institute of Physics, University of Belgrade \\
                      Pregrevica 118, 11080 Belgrade, Serbia}
\date{\today}
\maketitle

\begin{abstract}
Starting from the generalized pp waves that are exact vacuum solutions of
general relativity with a cosmological constant, we construct a new family
of exact vacuum solutions of Poincar\'e gauge theory, the generalized pp
waves with torsion. The ansatz for torsion is chosen in accordance with
the wave nature of the solutions. For a subfamily of these solutions, the
metric is dynamically determined by the torsion.
\end{abstract}

\section{Introduction}
\setcounter{equation}{0}

The principle of gauge symmetry was born in the work of Weyl \cite{x1},
where he obtained the electromagnetic field by assuming local $U(1)$
invariance of the Dirac Lagrangian. Three decades later, the Poincar\'e
gauge theory (PGT) was formulated by Kibble and Sciama \cite{x2}; it is
nowadays  a well-established gauge approach to gravity, representing a
natural extension of general relativity (GR) to the gauge theory of the
Poincar\'e group \cite{x3,x4}. Basic dynamical variables in PGT are the
tetrad field $b^i$ and the Lorentz connection $\om^{ij}=-\om^{ji}$
(1-forms), and the associated field strengths are the torsion $T^i=d
b^i+\om^i{_k}\wedge b^k$ and the curvature
$R^{ij}=d\om^{ij}+\om^i{_k}\wedge\om^{kj}$ (2-forms). By construction, PGT
is characterized by a Riemann-Cartan geometry of spacetime, and its
physical content is directly related to the existence of mass and spin as
basic characteristics of matter at the microscopic level. An up-to-date
status of PGT can be found in a recent reader with reprints and comments
\cite{x5}.

General PGT Lagrangian $L_G$ is at most quadratic in the field strengths.
The number of independent (parity invariant) terms in $L_G$ is nine, which
makes the corresponding dynamical structure rather complicated.  As is
well known from the studies of GR, exact solutions have an essential role
in revealing and understanding basic features of the gravitational
dynamics \cite{x6,x7,x8,x9}. This is also true for PGT, where exact
solutions allow us, among other things, to study the interplay between
dynamical and geometric aspects of torsion \cite{x5}.

In the context of GR, one of the best known families of exact solutions is
the family of \emph{pp waves}: it describes plane fronted waves with
parallel rays propagating on the Minkowski background $M_4$, see, for
instance, Ehlers and Kundt \cite{x6}. There is an important generalization
of this family, consisting of the exact vacuum solutions of GR with a
cosmological constant (\grl) such that for $\L\to 0$, they reduce to the
pp waves in $M_4$. We will refer to this family as the generalized pp
waves, or just \emph{\ppl\ waves} for short. The family of \ppl\ waves
belongs to a more general family, known as the Kundt class of type N,
labeled \kn{(\L)}; details on the \kn{(\L)} spacetimes can be found in the
monograph by Griffiths and Podolsk\'y \cite{x9}, see also Refs.
\cite{x10,x11,x12}. In this paper, we start from the Riemannian \ppl\
waves in \grl\ and construct a new family of the \emph{\ppl\ waves with
torsion}, representing a new class of exact vacuum solutions of PGT. The
torsion is introduced relying on the approach used in our previous paper
\cite{x13}. The present work is motivated by earlier studies of the exact
wave solutions in PGT \cite{x14}, and is regarded as a complement to them.

The paper is organized as follows. In section 2, we give a short account
of the Riemannian \ppl\ waves, including the relevant geometric and
dynamical aspects, as a basis for their extension to \ppl\ waves with
torsion. In section 3, we first introduce an ansatz for the new,
Riemann--Cartan (RC) connection, the structure of which complies with the
wave nature of a RC spacetime. The ansatz is parametrized by a specific
1-form $K$ living on the wave surface, and the related torsion has only
one, tensorial irreducible component. Then, we use the PGT field equations
to show that the dynamical content of K is described by two torsion modes
with the spin-parity values $J^P=2^+$ and $2^-$. In section 4, we find
solutions for both the metric function $H$ and the torsion function $K$,
in the spin-$2^+$ sector and for $\l>0,<0$ and $=0$. It is shown that $K$
has a decisive influence on the solution for $H$, and consequently, on the
resulting metric. In section 5, we shortly discuss solutions in the
spin-$2^-$ sector, which are found to be much less interesting. Section 6
concludes the exposition with a few remarks on some issues not covered in
the main text, and Appendices are devoted to certain technical details.

Our conventions are as follows. The Latin indices $(i, j, ...)$ refer to
the local Lorentz (co)frame and run over $(0,1,2,3)$, $b^i$ is the tetrad
(1-form), $h_i$ is the dual basis (frame), such that $h_i\inn b^k=\d^i_k$;
the volume 4-form is $\heps=b^0\wedge b^1\wedge b^2\wedge b^3$, the Hodge
dual of a form $\a$ is $\hd\a$, with $\hd 1=\heps$, totally antisymmetric
tensor is defined by $\hd(b_i\wedge b_j\wedge b_k\wedge b_m)=\ve_{ijkm}$
and normalized to $\ve_{0123}=+1$; the exterior product of forms is
implicit, except in Appendix \ref{appB}.

\section{Riemannian pp\mb{_\Lambda} waves}
\setcounter{equation}{0}

In this section, we give an overview of Riemannian \ppl\ waves using
the tetrad formalism \cite{x15}, necessary for the transition to PGT.

\subsection{Geometry}

The family of \ppl\ waves is a specific subclass of the Kundt spacetimes
\kn{(\L)}, labeled by \kn{(\l)[\a=1,\b=0]}; for the full classification of
the \kn{(\L)} spacetimes, see Refs. \cite{x9,x10}. In suitable local
coordinates $x^\m=(u,v,y,z)$, see Appendix \ref{appA}, the metric of the
\ppl\ waves can be written as
\bsubeq\lab{2.1}
\be
ds^2=2\left(\frac{q}{p}\right)^2du(Sdu+dv)
                              -\frac{1}{p^2}(dy^2+dz^2)\, ,     \lab{2.1a}
\ee
where
\be
p=1+\frac{\l}{4}(y^2+z^2)\, ,\qquad q=1-\frac{\l}{4}(y^2+z^2)\,,\qquad
S=-\frac{\l}{2}v^2+\frac{p}{2q}H(u,y,z)\, ,
\ee
\esubeq
with $\l$ being a suitably normalized cosmological constant, and the
unknown metric function $H$ is to be determined by the field equations.
The coordinate $v$ is an affine parameter along the null geodesics $x^\m=
x^\m(v)$, and $u$ is retarded time such that $u=$ const. are the spacelike
surfaces parametrized by $x^\a=(y,z)$. Since the null vector
$\xi=\xi(u)\pd_v$ is orthogonal to these surfaces, they are regarded as
wave surfaces, and $\xi$ is the null direction (ray) of the wave
propagation. The vector $\xi$ is not covariantly constant, and
consequently, the wave rays are not parallel and the wave surfaces are not
flat. For $\l\to 0$, the metric \eq{2.1} reduces to the metric of pp waves
on the $M_4$ background, which explains the term generalized pp waves, or
\ppl\ waves.

Next, we choose the tetrad field (coframe) in the form
\bsubeq\lab{2.2}
\be
b^0:=du\, ,\qquad b^1:=\left(\frac{q}{p}\right)^2(Sdu+dv)\, ,\qquad
b^2:=\frac{1}{p}dy\, ,\qquad b^3:=\frac{1}{p}dz\, ,
\ee
so that $ds^2=\eta_{ij}b^i\otimes b^j$, where $\eta_{ij}$ is the half-null
Minkowski metric:
$$
\eta_{ij}=\left( \ba{cccc}
             0 & 1 & 0 & 0  \\
             1 & 0 & 0 & 0  \\
             0 & 0 & -1& 0  \\
             0 & 0 & 0 & -1
                \ea
       \right)\, .
$$
The corresponding dual frame $h_i$ is given by
\be
h_0=\pd_u-S\pd_v\, ,\qquad h_1=\left(\frac{p}{q}\right)^2\pd_v\, ,
\qquad h_2=p\pd_y\, ,\qquad h_3=p\pd_z\, .
\ee
\esubeq
For the coordinates $x^\a=(y,z)$ on the wave surface, we have:
$$
x^c=b^c{_\a}x^\a=\frac{1}{p}(y,z)\, ,\qquad
\pd_c=h_c{^\a}\pd_\a=p(\pd_y,\pd_z)\, ,
$$
where $c=2,3$.

Starting from the general formula for the Riemannian connection 1-form,
$$
\om^{ij}:=-\frac{1}{2}\Bigl[h^i\inn db^j-h^j\inn db^i
                           -(h^i\inn h^j\inn db^k)b_k\Bigr]\, ,
$$
one can find its explicit form; for $i<j$, it reads:
\bsubeq\lab{2.3}
\bea
&&\om^{01}=\l v b^0-\frac{1}{q}(\l y b^2+\l z b^3)\,,
  \qquad \om^{02}=\frac{\l y}{q}b^0\, ,
  \qquad \om^{03}=\frac{\l z}{q}b^0\, ,                         \nn\\
&&\om^{12}=\frac{\l y}{q}b^1-\frac{q^2}{p}\pd_y S\,b^0\,,\qquad
  \om^{13}=\frac{\l z}{q}b^1-\frac{q^2}{p}\pd_z S\,b^0\,,\nn\\
&&\om^{23}=\frac{1}{2}(\l zb^2-\l yb^3)\, ,
\eea
Introducing the notation $i=(A,a)$, where $A=0,1$ and $a=(2,3)$, one can
rewrite $\om^{ij}$ in a more compact form:
\bea
&&\om^{01}=\l v b^1-\frac{2}{qp}(b^c\pd_c p)\,,\nn\\
&&\om^{Ac}=-\frac{2}{qp}b^A\pd^c p
           +k^A\frac{q^2}{p^2}b^0\pd^c S\, ,                    \nn\\
&&\om^{23}=-\frac{1}{p}(b^2\pd^3 p-b^3\pd^2 p)\, ,
\eea
\esubeq
where $k^i=(0,1,0,0)$ is a null propagation vector, $k^2=0$.

The above connection defines the Riemannian curvature
$R^{ij}=d\om^{ij}+\om^i{_m}\om^{mj}$; for $i<j$, it is given by
\bsubeq\lab{2.4}
\be
R^{ij}=\left\{\ba{rl}
              -\l b^1b^c+k^1b^0Q^c,&\text{ for } (i,j)=(1,c) \\[3pt]
              -\l b^ib^j,& \text{ otherwise},
              \ea\right.
\ee
where $Q^c$ is a 1-form introduced by Obukhov \cite{x15},
\be
Q_c=-\ul{\nab}\left[\left(\frac{q}{p}\right)^2 h_c\inn\ul{d} S\right]
    +\left(\frac{q}{p}\right)^3h_c\inn\left[
       \ul{d}\left(\frac{p}{q}\right)\wedge\ul{d}S\right]\, ,   \nn
\ee
and $\ul{d}=dx^\a\pd_\a$ is the exterior derivative on the wave surface.
In more details:
\bea
&&Q^2=\frac{q}{2p}\left[2p^2\pd_y\left(\frac{q}{p}\pd_y S\right)
                          +q\l y\pd_y S-q\l z\pd_z S\right]b^2
     +\frac{q}{2}\left[2q\pd_{yz}S-\l z\pd_y S
                                  -\l y\pd_z S\right]b^3,       \nn\\
&&Q^3=\frac{q}{2p}\Bigl[2p^2\pd_z\left(\frac{q}{p}\pd_z S\right)
                          +q\l z\pd_z S-q\l y\pd_yS\Bigr]b^3
 +\frac{q}{2}\Bigl[2q\pd_{yz}S-\l z\pd_y S-\l y\pd_z S\Bigr]b^2.\nn
\eea
As a consequence, $R^{ij}$ can be represented more compactly as
\be
R^{ij}=-\l b^ib^j+2b^0k^{[i}Q^{j]}\, .
\ee
\esubeq
The Ricci 1-form $\ric^i:=h_m\inn\ric^{mi}$ is expressed in terms of
$Q=h_c\inn Q^c$,
\bea
&&\ric^i=-3\l b^i+b^0k^i Q\, ,                                  \nn\\
&&Q=qp\left[\pd_y\left(\frac{q}{p}\pd_y S\right)
             +\pd_z\left(\frac{q}{p}\pd_z S\right)\right]
  =\frac{qp}{2}\left[\pd_{yy}H+\pd_{zz}H
                              +\frac{2\l}{p^2}H\right],\quad \lab{2.5}
\eea
and the scalar curvature $R:=h_i\inn\ric^i$ reads:
\be
R=-12\l\, .
\ee

\subsection{Dynamics}

\subsubsection*{pp\mb{_\Lambda} waves in GR\mb{_\L}}

Starting with the action $I_0=-\int d^4 x(a_0R+2\L_0)$, one can derive the
\grl\ field equations in vacuum:
\bsubeq\lab{2.6}
\be
2a_0G^n{_i}-2\L_0\d^n_i=0\, ,
\ee
where $G^n{_i}$ is the Einstein tensor. The trace and the traceless piece
of these equations read:
\be
\L_0=3a_0\l\, ,\qquad
\ric^i-\frac{1}{4}Rb^i\equiv b^0k^i Q=0\,  .
\ee
\esubeq
As a consequence, the metric function $H$ must obey
\be
\pd_{yy}H+\pd_{zz}H+\frac{2\l}{p^2}H=0\, .                      \lab{2.8}
\ee
There is a simple solution of these equations,
\be
H_c=\frac{1}{p}\bigl(A(u)q+B_\a x^\a\bigr)f(u)\, ,
\ee
for which $Q^a$ vanishes. This solution is trivial (or pure gauge), since
the associated curvature takes the background form, $R^{ij}=-\l b^ib^j$;
moreover, it is conformally flat, since its Weyl curvature vanishes. The
nontrivial vacuum solutions are characterized by $Q=0$, but $Q^c\ne 0$;
their general form can be found in \cite{x10}.

\subsubsection*{pp\mb{_\Lambda} waves in PGT}

To better understand the relation between \grl\ and PGT, it is interesting
to examine whether \ppl\ waves satisfying the \grl\ field equations in
vacuum are also vacuum solution of PGT. It turns out that a more general
version of the problem has been already solved by Obukhov \cite{x4}.
Studying the PGT field equations for torsion-free configurations, he
proved the following important theorem:
\bitem
\item[T1.] In the absence of matter, any solution of \grl\ is a
    torsion-free solution of PGT.
\eitem
It is interesting to note that the inverse statement, that any
torsion-free vacuum solution of PGT is also a vacuum solution of \grl, is
also true, except for three specific choices of the PGT coupling
constants.

\section{pp\mb{_\Lambda} waves with torsion}\label{sec3}
\setcounter{equation}{0}

In this section, we extend the \ppl\ waves that are vacuum solutions of
\grl\ to a new family of the exact vacuum solutions of PGT, characterized
by the existence of torsion.

\subsection{Ansatz}

The main step in constructing the \ppl\ waves with torsion is to find an
ansatz for torsion that is compatible with the wave nature of the
solutions. This is achieved by introducing torsion at the level of
connection.

Looking at the Riemannian connection \eq{2.3}, one can notice that its
radiation piece appears only in the $\om^{1c}$ components:
$$
(\om^{1c})^R=\frac{q^2}{p^2}(h^{c\a}\pd_\a S) b^0\, .
$$
This motivates us to construct new connection by applying the rule
\bsubeq\lab{3.1}
\be
\pd_\a S\to \pd_\a S+K_\a\,, \qquad K_\a=K_\a(u,y,z)\, ,
\ee
where $K_\a$ is the component of the 1-form $K=K_\a dx^\a$ on the wave
surface. Thus, the new form of $(\om^{ij})^R$ reads
\be
(\om^{ic})^R:=k^i\frac{q^2}{p^2}h^{c\a}(\pd_\a S+K_\a)\,b^0\, ,
\ee
\esubeq
whereas all the other non-radiation pieces retain their Riemannian form
\eq{2.3}.

The geometric content of the new connection is found by calculating the
torsion:
\be
T^i=\nab b^i+\om^i{_m}b^m=k^i\frac{q^2}{p}b^0(b^2 K_y+b^3K_z)
   =k^i\frac{q^2}{p^2}b^0b^cK_c\, .                             \lab{3.2}
\ee
The only nonvanishing irreducible piece of $T^i$ is ${}^{(1)}T^i$.

The new connection modifies also the curvature, so that its radiation
piece becomes
\bsubeq
\be
(R^{1c})^R=k^1b^0(Q^c+\Th^c)\, ,
\ee
where
\bea
&&\Th^2=\frac{q}{2p}\Bigl[\bigl(2qp\pd_y K_y-pK_y\l y-qK_z\l z\bigr)b^2
        +\bigl(-2qp\pd_z K_y+pK_y\l z-qK_z\l y\bigr)b^3\Bigr]\,,\nn\\
&&\Th^3=\frac{q}{2p}\Bigl[\bigl(2qp\pd_z K_z-pK_z\l z-qK_y\l y\bigr)b^3
        +\bigl(-2qp\pd_y K_z+pK_z\l y-qK_y\l z\bigr)b^2\Bigr].\nn
\eea
The covariant form of the curvature reads
\be
R^{ij}=-\l b^ib^j+2b^0k^{[i}(Q^{j]}+\Th^{j]})\, ,
\ee
and the Ricci curvature takes the form
\be
\ric^i=-3\l b^i+b^0k^i(Q+\Th)\, ,
\ee
where $\Th:=h_c\inn\Th^c$. The torsion has no influence on the scalar
curvature:
\be
R=-12\l\, .
\ee
\esubeq

Thus, our ansatz defines a RC geometry of spacetime.

\subsection{PGT field equations}\label{sub32}

Having adopted the ansatz for torsion defined in Eq. \eq{3.1}, we now wish
to find explicit form of the PGT field equations and use them to determine
dynamical content of our ansatz.

As shown in Appendices \ref{appB} and \ref{appC}, the field equations
depend on the structure of the irreducible components of the field
strengths. For torsion, we already know that the only nonvanishing
irreducible component is ${}^{(1)}T_i=T_i$, defined in Eq. \eq{3.2}. As
for the curvature, we note that our ansatz yields $X=0$ and $b^m\ric_m=0$,
where $X$ is defined in \eq{B.2b}. Then, the irreducible decomposition of
the curvature given in \eq{B.2a} implies
\be
{}^{(3)}R_{ij}=0\, ,\qquad {}^{(5)}R_{ij}=0\, ,
\ee
whereas the remaining pieces ${}^{(n)}R_{ij}$ are given in terms of the
1-forms
\bea\lab{3.5}
&&\Phi^i=k^i b^0(Q+\Th)\, ,\qquad
  \Th=qp\left[\pd_y\left(\frac{q}{p}K_y\right)
             +\pd_z\left(\frac{q}{p}K_z\right)\right]\, ,      \nn\\                                       \nn\\
&&\Psi^i=X^i=-k^i b^0\S\, ,\qquad
  \S=qp\left[\pd_z\left(\frac{q}{p}K_y\right)
            -\pd_y\left(\frac{q}{p}K_z\right)\right] \, .
\eea
After calculating ${}^{(1)}T_i$ and ${}^{(n)}R_{ij}$, the procedure
described in Appendix \ref{appC} leads to the following form of the two
PGT field equations \eq{C.3}:
\bsubeq\lab{3.6}
\bea
\text{(1ST)}
&&\L_0=3a_0\l\, ,\qquad a_1\Th-A_0(Q+\Th)=0\, ,                 \lab{3.6a}\\[3pt]
\text{(2ND)}
&&-(b_2+b_1)(\nab\Psi^1)b^2-(b_4+b_1)(\nab\Phi^1)b^3
                 -2\bigl(a_0-A_1\bigr)T^1b^3=0\, ,\quad         \nn\\
&&-(b_2+b_1)(\nab\Psi^1)b^3+(b_4+b_1)(\nab\Phi^1)b^2
                 +2\bigl(a_0-A_1\bigr)T^1b^2=0\, ,\quad         \lab{3.6b}
\eea
\esubeq
where $A_0=a_0+(b_4+b_6)\l$ and $A_1=a_1-(b_6-b_1)\l$~ \cite{x16}.

Leaving (1ST) as is, (2ND) can be given a more clear structure as follows:
\bitem
\item[$-$] use (1ST) to express $\Phi^1=b^0(Q+\Th)$ in the form
    $\Phi^1=(a_1/A_0)b^0\Th$;\vsm
\item[$-$] multiply (2ND) by $A_0/q$.
\eitem
As a result, the previous two components of (2ND) transform into:
\bsubeq\lab{3.7}
\bea
&&A_0(b_2+b_1)\pd_z(p\S/q)+a_1(b_4+b_1)\pd_y(p\Th/q)
                 +2A_0(a_1-A_0)(q/p)K_y=0\, ,  \qquad           \lab{3.7a}\\[3pt]
&&-A_0(b_2+b_1)\pd_y(p\S/q)+a_1(b_4+b_1)\pd_z(p\Th/q)
                 +2A_0(a_1-A_0)(q/p)K_z=0\, .   \qquad          \lab{3.7b}
\eea
\esubeq
Then, calculating $\pd_y$\eq{3.7a}$+\pd_z$\eq{3.7b} and
$\pd_z$\eq{3.7a}$-\pd_y$\eq{3.7b} yields the final form of (2ND):
\bsubeq\lab{3.8}
\bea
&&(\pd_{yy}+\pd_{zz})(p\Th/q)
                   -m_{2^+}^2\dis\frac{1}{p^2}(p\Th/q)=0\,,
       \quad\dis m_{2^+}^2:=\frac{2A_0(a_0-A_1)}{a_1(b_1+b_4)},  \lab{3.8a}\\[3pt]
&&(\pd_{yy}+\pd_{zz})(p\S/q)
                   -m_{2^-}^2\dis\frac{1}{p^2}(p\S/q)=0\,,
       \quad\dis m_{2^-}^2:=\frac{2(a_0-A_1)}{b_1+b_2}.          \lab{3.8b}
\eea
\esubeq
The parameters $m^2_{2^\pm}$ have a simple physical interpretation. In the
limit $\l\to 0$, they represent masses of the spin-$2^\pm$ torsion modes
with respect to the $M_4$ background \cite{x17},
$$
\bar m^2_{2^+}=\frac{2a_0(a_0-a_1)}{a_1(b_1+b_4)}\, ,\qquad
\bar m_{2^-}^2=\frac{2(a_0-a_1)}{b_1+b_2}\, ,
$$
whereas for finite $\l$, $m^2_{2^\pm}$ are associated to the torsion modes
with respect to the (A)dS background.

In $M_4$, the physical torsion modes are required to satisfy the
conditions of no ghosts (positive energy) and no tachyons (positive $m^2$)
\cite{x17,x18}. However, for spin-$2^+$ and spin-$2^-$ modes, the
requirements for the absence of ghosts, given by the conditions
$b_1+b_2<0$ and $b_1+b_4>0$, do not allow for both $m^2$ to be positive.
Hence, only one of the two modes can exist as a propagating mode (with
finite mass), whereas the other one must be ``frozen" (infinite mass).
Although these conditions refer to the $M_4$ background, we assume their
validity also for the (A)dS background, in order to have a smooth limit
when $\l\to 0$.

One should note that the two spin-$2$ sectors have quite different
dynamical structures.
\bitem
\item[$-$] In the spin-$2^-$ sector, the infinite mass of the spin-$2^+$
    mode implies $\Th=0$, whereupon (1ST) yields $Q=0$, which is exactly
    the \grl\ field equation for metric. Thus, the existence of torsion
    has no influence on the metric.\vsm
\item[$-$] In the spin-$2^+$ sector, the infinite mass of the spin-$2^-$
    mode implies $\S=0$, whereas (1ST) yields that $Q$ is proportional to
    $\Th$, with $\Th\ne 0$. Thus, the torsion function $\Th$ has a
    decisive dynamical influence on the form of metric.
\eitem

In the next section, we focus our attention on the spin-$2^+$ sector,
where the metric appears to be a genuine dynamical effect of PGT.

\section{Solutions in the spin-$2^+$ sector}
\setcounter{equation}{0}

In this section, we first find solutions of Eq. \eq{3.8a} for the
spin-$2^+$ mode $V=(p/q)\Th$, and then use that $V$ to find the metric
function $H$ and the torsion functions $K_\a$, the quantities that
completely define geometry of the \ppl\ waves with torsion.

\subsection{Solutions for \mb{V=(p/q)\Th}}\label{sub41}

The field equation for the spin-$2^+$ sector can be written in a slightly
simpler form as
\be
(\pd_{yy}+\pd_{zz})V-\frac{m^2}{p^2}V=0\, ,                     \lab{4.1}
\ee
where $V=(p/q)\Th$ and $m^2=m^2_{2^+}$. We have seen in Appendix
\ref{appA} that local coordinates $(y,z)$ are well-defined in the region
where $p$ and $q$ do not vanish,  which is an open disk of finite radius,
$y^2+z^2<4|\l|^{-1}$. Since \eq{4.1} is a differential equation with
circular symmetry, it is convenient to introduce polar coordinates,
$y=\r\cos\vphi,z=\r\sin\vphi$, in which Eq. \eq{4.1} takes the form
\bsubeq\lab{4.2}
\be
\left(\frac{\pd^2}{\pd\r^2}+\frac{1}{\r}\frac{\pd}{\pd \r}
+\frac{1}{\r^2}\frac{\pd^2}{\pd\vphi^2}\right)V-\frac{m^2}{p^2}V=0\, .
\ee
Looking for a solution of $V$ in the form of a Fourier expansion,
$$
V=\sum_{n=0}^\infty V_n(\r)(c_ne^{in\vphi}+\bar c_n e^{-in\vphi})\, ,
$$
we obtain:
\be
V_n''+\frac{1}{\r}V_n'
     -\left(\frac{n^2}{\r^2}+\frac{m^2}{p^2}\right)V_n=0\, ,    \lab{4.2b}
\ee
\esubeq
where prime denotes $d/d\r$.

\subsubsection*{\mb{\l/4\equiv\ell^{-2}>0}}

Let us first consider solutions of the dS type, using a convenient
notation:
$$
x=\frac{\r}{\ell},\quad \m=m\ell,\qquad
   \xi=\frac{1}{2}\left(1+\sqrt{1-\m^2}\right)\, .\quad
$$
The general solutions of \eq{4.2b} for $n=0$ and $n>0$ are given by:
\bsubeq\lab{4.3}
\bea
V_0&=&c_1(1+x^2)^{1-\xi}\,
      {}_2F_1\bigl(1-\xi,1-\xi;2(1-\xi);-|1+x^2|\bigr)          \nn\\
  &&+c_2(1+x^2)^{\xi}\,{}_2F_1\bigl(\xi,\xi;2\xi;-|1+x^2|\bigr),\\
V_n&=&c_1(x^2)^{n/2}(1+x^2)^\xi\,
      {}_2F_1\bigl(\xi,\xi+n;1+n,-x^2\bigr)                     \nn\\
  &&+c_2(x^2)^{-n/2}(1+x^2)^\xi\,
      {}_2F_1\bigl(\xi,\xi-n;1-n,-x^2\bigr)\, ,
\eea
\esubeq
where $c_n=c_n(u)~ (n=1,2)$ and ${}_2F_1(a,b,c,z)$ is the hypergeometric
function \cite{x19}.

\subsubsection*{\mb{\l/4\equiv-\ell^{-2}<0}}

In the AdS sector, using
$$
\bar\xi=\frac{1}{2}\left(1+\sqrt{1+\m^2}\right)\, ,
$$
the solutions for $n=0$ and $n>0$ take the forms:
\bsubeq\lab{4.4}
\bea
V_0&=&c_1(1-x^2)^{1-\bar\xi}\,
      {}_2F_1\bigl(1-\bar\xi,1-\bar\xi;2(1-\bar\xi);|1-x^2|\bigr) \nn\\
  &&+c_2(1-x^2)^{\bar\xi}\,
    {}_2F_1\bigl(\bar\xi,\bar\xi;2\bar\xi;|1-x^2|\bigr)\, ,     \lab{4.4a}\\
V_n&=&c_1(x^2)^{n/2}(x^2-1)^{\bar\xi}\,
     {}_2F_1(\bar\xi,\bar\xi+n;1+n,x^2)                         \nn\\
  &&c_2(x^2)^{-n/2}(x^2-1)^{\bar\xi}\,
     {}_2F_1(\bar\xi,\bar\xi-n;1-n,x^2)\, .
\eea
\esubeq
These solutions are essentially an analytic continuation by $\ell\to
i\ell$ of those in Eq. \eq{4.3}.

\subsubsection*{\mb{\l=0}}

The general solution of Eq. \eq{4.2b} has the form
\be
V_n=c_1J_n(-im\r)+c_2Y_n(-im\r)\, ,\qquad n=0,1,2,\dots  \lab{4.5}
\ee
where $J_n$ and $Y_n$ are Bessel functions of the 1st and 2nd kind,
respectively.

\subsection{Solutions for the metric function \mb{H}}

Fora a given $\Th$, the first PGT field equation
$A_0Q=(a_1-A_0)\Th$, with $Q$ defined in \eq{2.5}, represents a
differential equation for the metric function $H$:
\be
(\pd_{yy}+\pd_{zz})H+\frac{2\l}{p^2}H
                       =\frac{2(a_1-A_0)}{A_0}\frac{1}{p^2}V\, .\lab{4.6}
\ee
This is a second order, linear nonhomogeneous differential equation, and
its general solution can be written as
$$
H=H^h+H^p\, ,
$$
where $H^h$ is the general solution of the homogeneous equation, and $H^p$
a particular solution of \eq{4.6}. By comparing Eq. \eq{4.6} to Eq.
\eq{4.1}, one finds a simple particular solution for $H$:
\bsubeq\lab{4.7}
\be
H^p=\s V\, ,\qquad \s=\frac{2(a_1-A_0)}{(2\l+m^2)A_0}\, .
\ee
On the other hand, $H^h$ coincides with the general vacuum solution of
\grl, see \eq{2.8}. Since our idea is to focus on the genuine torsion
effect on the metric, we choose $H^h=0$ and adopt $H^p$ as the most
interesting PGT solution for the metric function $H$. Thus, we have
\be
H_n=\s V_n\, .
\ee
\esubeq

\subsection{Solutions for the torsion functions \mb{K_\a}}

In the spin-$2^+$ sector, the torsion functions $K_\a$ can be determined
from Eqs. \eq{3.7}, combined with the condition $\S=0$:
\be
\pd_y V+m^2\frac{q}{p}K_y=0\,,\qquad
\pd_z V+m^2\frac{q}{p}K_z=0\, .                                 \lab{4.8}
\ee
Going over to polar coordinates,
$$
K_y=K_\r\cos\vphi-\frac{K_\vphi}{\r}\sin\vphi\, ,\qquad
K_z=K_\r\sin\vphi+\frac{K_\vphi}{\r}\cos\vphi\, ,
$$
the previous equations are transformed
into
\bsubeq
\be
K_\r=-\frac{1}{m^2}\frac{p}{q}\pd_\r V\, ,\qquad
K_\vphi=-\frac{1}{m^2}\frac{p}{q}\pd_\vphi V\, ,
\ee
or equivalently, in terms of the Fourier modes,
\be
K_{\r n}=-\frac{1}{m^2}\frac{p}{q}\pd_\r V_n\, ,\qquad K_{\vphi
n}=-\frac{1}{m^2}\frac{p}{q}nV_n\, ,
\ee
\esubeq
where $K_\vphi=\sum_{n=1}^\infty (d_n e^{in\vphi}+\bar d_n e^{-i n\vphi})$
with $d_n=-ic_n$, and similarly for $K_\r$.

\subsection{Graphical illustrations}

Here, we illustrate graphical forms of two specific solutions by giving
plots of their metric functions $H$ and the typical torsion component
$T^1{}_{02}$,
\bea
&&H=\s V\, ,                                                    \nn\\
&&T^1{}_{02}=\frac{q^2}{p^2}K_2=\frac{q^2}{p}K_y
  =-\frac{1}{m^2}q\bigl(\pd_\r V\cos\vphi-\r^{-1}K_\vphi\sin\vphi\bigr).
\eea
For $\l\ne 0$, it is convenient to use the units in which $\ell=1$.
\begin{figure}[ht]
\centering
\includegraphics[height=4cm]{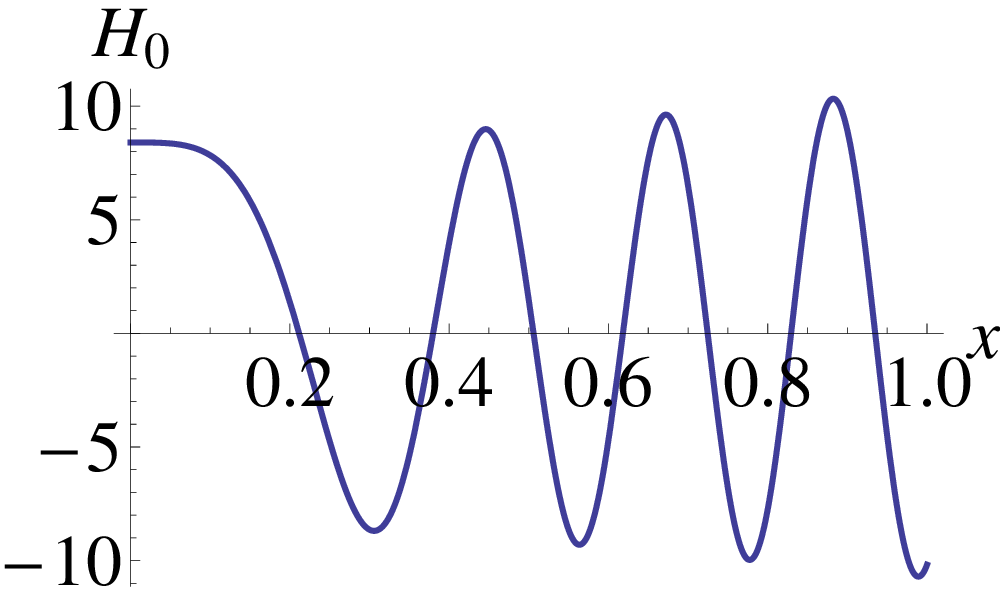}\qquad
\includegraphics[height=4cm]{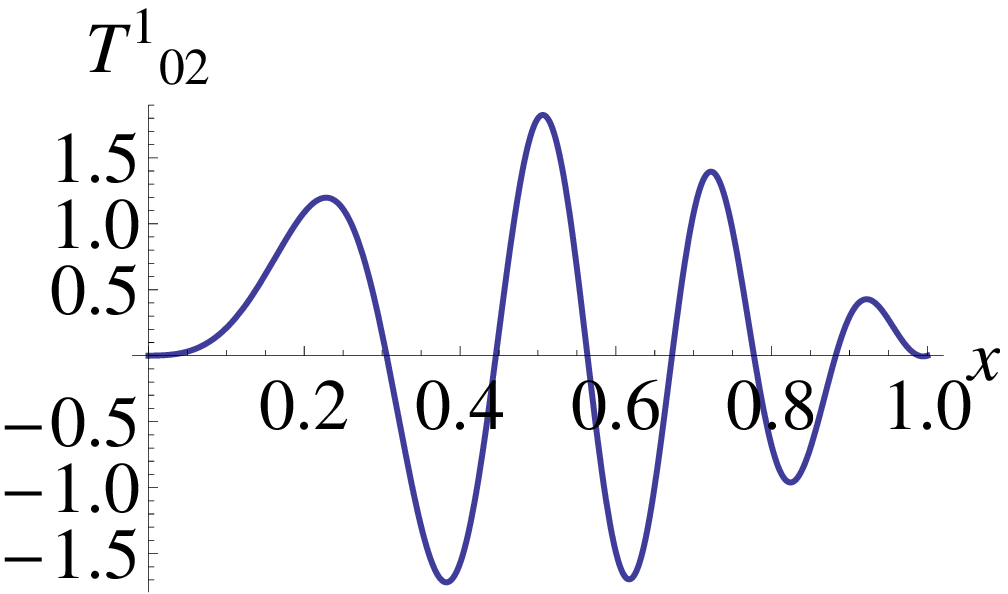}
\caption{The plots of a solution in the sector $\l>0$, in units $\ell=1$,}
{for $n=0,\m=100,c_1=1,c_2=0,\s=1$;~~ left is~  $H_0$,~ right is~
$T^1{}_{02}(\vphi=0)$.}\label{fig1}
\end{figure}
\begin{figure}[ht]
\centering
\includegraphics[height=4cm]{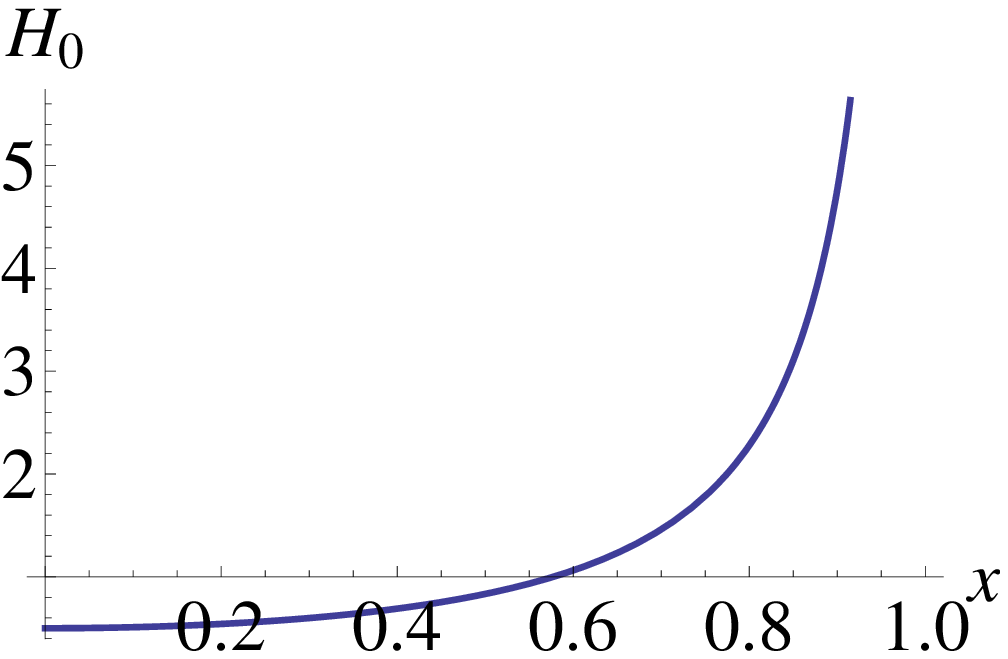}\qquad
\includegraphics[height=4cm]{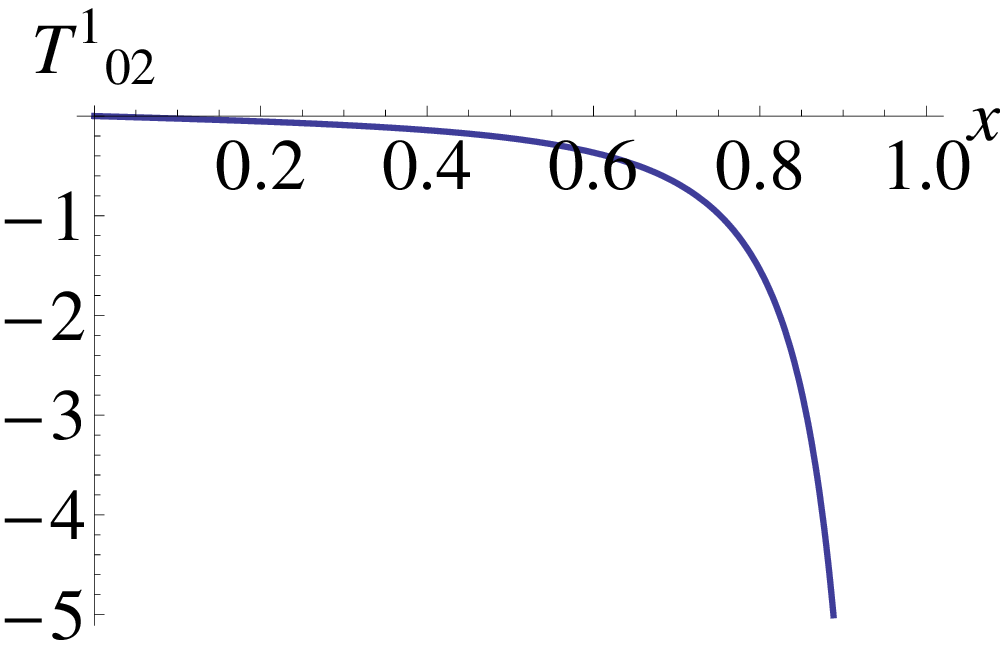}
\caption{The plots of a solution in the sector $\l<0$, in units $\ell=1$,}
{for $n=0,\m=\sqrt{8},c_1=0.1,c_2=0$;~~ left is~ $H_0$,~ right is~
$T^1{}_{02}(\vphi=0)$.}\label{fig2}
\end{figure}

In the dS sector (Figure 1), the zero modes of both $H$ and
$T^1{}_{02}(\vphi=0)$ are regular functions with a clear-cut wave-like
behavior in the region $0<x<1$. The plots correspond to the \ppl\ geometry
for fixed $u$, and as $u$ increases, the pictures change. In the AdS
sector (Figure 2), the solution is singular at $x=1$, or equivalently at
$p=0$, and it does not have a typical wave-like shape. For a discussion of
the singularity at $p=0$, see Ref. \cite{x11}. We also examined a zero
mode solution ($n=0$) in the $M_4$ sector ($\l=0$); its shape is similar
to what we have in Figure 2, but it remains finite at $x=1$.

\section{Solutions in the spin-$2^-$ sector}
\setcounter{equation}{0}

As we noted at the end of section \ref{sec3}, the spin-$2^-$ sector is
characterized by $\Th=0$ and, as a consequence of (1ST), by $Q=0$.
Equation \eq{3.8b} for $\S$ reads
\be
(\pd_{yy}+\pd_{zz})U-\frac{m^2}{p^2}U=0\, ,
\ee
where $U=(p/q)\S$ and $m^2=m_{2^-}^2$. Clearly, the solutions for $U$
coincide with the solutions for $V=(p/q)\Th$ in subsection \ref{sub41}.
Furthermore, the metric function $H$, defined by $Q=0$, has the \grl\
form, and the solutions for the torsion functions $K_\a$ follow from the
two equations
\be
\pd_y U+m^2\frac{q}{p}K_y=0\, ,\qquad \pd_z U+m^2\frac{q}{p}K_z=0\, ,
\ee
the counterparts of those in \eq{4.8}.

The fact that the metric of the spin-$2^-$ sector is independent of
torsion makes this sector, in general, much less interesting. There is,
however, one solution in this sector that should be mentioned: it is the
solution with $H=0$ for which the metric takes the (A)dS/$M_4$ form, and
the complete dynamics is carried solely by the torsion. We skip discussing
details of this case, as they can be easily reconstructed from the results
given in the previous section, following the procedure outlined above.

\section{Concluding remarks}
\setcounter{equation}{0}

In this paper, we found a new family of the exact vacuum solutions of PGT,
the family of the \ppl\ waves with torsion. Here, we wish to clarify a few
issues that have not been properly covered in the main text.

The essential step in our construction is the ansatz for the RC connection
\eq{3.1}, which modifies only the radiation piece of the corresponding
Riemannian connection \eq{2.3}. A characteristic feature of the resulting
solution is the presence of the null vector $k^i=(0,1,0,0)$ in the
spacetime geometry. The vector field $k^i\pd_i=(p/q)^2\pd_v$ is orthogonal
to the spatial surfaces $u$ = const., and is interpreted as the
propagation vector of the \ppl\ wave with torsion. Is such an
interpretation justifiable?

Although gravitational waves belong to one of the best known families of
exact solutions in \grl, a unique covariant criterion for their precise
identification is still missing. One of the early criteria of this type
was formulated by Lichnerowicz, based on an analogy with methods used to
determine electromagnetic radiation, see Zakharov \cite{x7}. This
criterion can be formulated as a requirement that the radiation piece of
the curvature, $S^{ij}=R^{ij}+\l b^ib^j$, satisfies the radiation
conditions:
\bsubeq
\be
k^iS_{ij}=0\, ,\qquad \ve^{ijkn}k_jS_{kn}=0\, .
\ee
However, when applied to a RC geometry, the Lichnerowicz criterion can be
naturally extended to include the torsion 2-form:
\be
k^i T_i=0\, ,\qquad \ve^{ijmn}k_mT_n=0\, .
\ee
\esubeq
A direct comparison to the expressions \eq{2.4} and \eq{3.2} shows that
both sets of the radiation conditions are satisfied. This result gives a
strong support to interpreting the \ppl\ waves with torsion as
proper wave solutions of PGT.

Looking at the explicit solutions for the \ppl\ waves with torsion, one
should note that, in general, the hypergeometric function
${}_2F_1(a,b,c,x)$ is singular at $x=1$ $(\r=\ell)$ \cite{x19}; moreover,
local coordinates we are using are singular at both $x=1$ and $x=0$
(Appendix \ref{appA}). To test the nature of these singularities, we
calculated the following torsion and curvature invariants:
\bea
&&T^i\wedge \hd T_i=0\, ,   \nn\\
&&R=-12\l\,,\qquad R^{ij}\wedge\hd R_{i,j}=12\l^2\heps\, ,\qquad
  R^{ij}{}_{kl}R^{kl}{}_{mn}R^{mn}{}_{ij}=-48\l^3\, ,           \lab{5.2}
\eea
the fourth order invariant is $96\l^4$, and so on. All these invariants
are well-behaved at $x=1,0$, which might be a signal that the
singularities in question are just the coordinate singularities. However,
according to Wald \cite{x20}, the geometric singularities are not always
visible in the field strength invariants. Hence, this issue deserves
further clarification.

If the curvature $R^{ij}$ is replaced by its radiation piece $S^{ij}$, all
the invariants in \eq{5.2} are found to vanish. According to Bell's second
criterion \cite{x7}, we have here another result that supports the wave
interpretation of our \ppl\ solutions.

\medskip
In conclusion, the family of solutions that we found reveals an unexpected
dynamical aspect of torsion. Namely, although torsion is introduced by a
minor modification of the Riemannian connection, see \eq{3.1}, the metric
function $H$ in \eq{4.7} is determined solely by the torsion, and
consequently, the form of the metric is a genuine dynamical effect of PGT.
A more detailed information could be obtained by analyzing the motion of
test particles/fields in the RC spacetimes associated to the \ppl\ waves
with torsion.

\section*{Acknowledgments}

One of us (M.B.) would like to thank Yuri Obukhov for the correspondence
and information on his work on the same subject. This work was supported
by the Serbian Science Foundation under Grant No. 171031.

\appendix
\section{On hyperbolic geometries}\label{appA}
\setcounter{equation}{0}

(A)dS space can be simply represented as a 4D hyperboloid $H_4$ embedded
in a 5D Minkowski space $M_5$ with metric $\eta_{MN}=(+,-,-,-,\s)$,
\bsubeq\lab{A.1}
\be
H_4:\quad X_0^2-X_1^2-X_2^2-X_3^2-\s X_5^2=-\s \ell^2\, ,
\ee
where $\s=+1$ for a dS space and $\s=-1$ for an AdS space \cite{x9,x21}. The
metric on $H_4$ reads
\be
ds^2=dX_0^2-dX_1^2-dX_2^2-dX_3^2-\s dX_5^2\, ,                  \lab{A.1b}
\ee
\esubeq
and its scalar curvature is $R=-12\s/\ell^2$. The group of isometries of
the dS/AdS spaces is $SO(1,4)/SO(2,3)$, and the corresponding topologies
are $R\times S^3$ for the dS space, and $S^1\times R^3$  for the AdS space
(or $R^4$ for its universal covering).

Going now back to the generalized pp wave metric \eq{2.1}, we note that in
the limit $H=0$, it  describes the background (A)dS geometry:
\bea
&&ds^2=2\left(\frac{q}{p}\right)^2du\left(-2\L v^2du+dv\right)
                                   -\frac{1}{p^2}(dy^2+dz^2)\, ,\lab{A.2}\\
&&p=1+\L(y^2+z^2),\qquad q=1-\L(y^2+z^2)\, .                    \nn
\eea
As we shall see below, $\L$ is related to $\ell$ by $4\s\L=1/\ell^2$;
moreover, $\L>0$ for dS and $\L<0$ for AdS. The two forms of the metric
associated to the hyperboloid $H_4$ are related to each other by a
coordinate transformation \cite{x11},
\bea
&&X_0=\frac{q}{2p}(u+v+\L u^2 v)\, ,\qquad~~
  u=2\s\ell\frac{ X_5-\sqrt{-\s(X_0^2-X_1^2-\s X_5^2)} }{X_0-X_1}\,,\nn\\[2pt]
&&X_1=\frac{q}{2p}(u-v+\L u^2 v)\, ,\qquad~~
  v=\frac{X_0-X_1}{4\ell\sqrt{-\s(X_0^2-X_1^2-\s X_5^2)}}\, ,   \nn\\[2pt]
&&X_2=\frac{y}{p}\, ,\quad X_3=\frac{z}{p}\, , \hspace{2cm}
  y=\frac{2\ell X_2}{\ell+\sqrt{\ell^2-\s(X_2^2+X_3^2)}}\, ,    \nn\\[2pt]
&&X_5=\frac{1}{2\sqrt{\s\L}}\frac{q}{p}(1+2\L uv),\hspace{0.9cm}
  z=\frac{2\ell X_3}{\ell+\sqrt{\ell^2-\s(X_2^2+X_3^2)}}\, .    \lab{A.3}
\eea
Indeed, the coordinates $X_M$ in $M_4$ describe the hyperboloid $H_4$,
$$
(X_0^2-X_1^2-\s X_5^2)-X_2^2-X_3^2=-\frac{1}{4\L}\frac{q^2}{p^2}
  -\frac{1}{p^2}(y^2+z^2)=-\frac{1}{4\L}=-\s\ell^2\, ,
$$
and the corresponding metric \eq{A.1b}, followed by the rescaling $v\to 2v$,
coincides with \eq{A.2}.

Since local coordinates $x^\m=(u,v,x,y)$ are introduced by the
parametrization \eq{A.3}, they are well defined for
$$
X_0^2-X_1^2-\s X_5^2=-\frac{1}{4\L}\frac{q^2}{p^2}>0.
$$
The limiting value $q=0$ is not allowed, as it represents the singularity
of the local coordinate system $(u,v,y,z)$; this singularity is visible
only for $\L>0$. The same conclusion follows from the fact that the
determinant of the metric \eq{A.2} vanishes for $q=0$. Furthermore, an
inspection of equations \eq{A.3} reveals the existence of another
singularity, located at $p=0$; it is visible only for $\L<0$. Thus, local
coordinates $(u,v,y,z)$ are restricted to the region where $q$ and/or $p$
do not vanish: $y^2+z^2\le |\L|^{-1}$. More on the geometric
interpretation of these singularities can be found in Ref. \cite{x11}.

\section{Irreducible decomposition of the field strengths}\label{appB}
\setcounter{equation}{0}

We present here formulas for the irreducible decomposition of the PGT
field strengths in a 4D Riemann--Cartan spacetime \cite{x4,x22}.

The torsion 2-form has three irreducible pieces:
\bea
&&{}^{(2)}T^i=\frac{1}{3}b^i\wedge(h_m\inn T^m)\, ,             \nn\\
&&{}^{(3)}T^i=\frac{1}{3}h^i\inn(T^m\wedge b_m)\, ,             \nn\\
&&{}^{(1)}T^i=T^i-{}^{(2)}T^i-{}^{(3)}T^i\, .
\eea
The RC curvature 2-form can be decomposed into six irreducible pieces:
\bsubeq
\be
\ba{ll}
{}^{(2)}R^{ij}={}^*(b^{[i}\wedge\Psi^{j]})\, ,
           &{}^{(4)}R^{ij}=b^{[i}\wedge\Phi^{j]}\, ,            \\[3pt]
{}^{(3)}R^{ij}=\dis\frac{1}{12}X\,{}^*(b^i\wedge b^j)\, ,
           &{}^{(6)}R^{ij}=\dis\frac{1}{12}F\,b^i\wedge b^j\, , \\[9pt]
{}^{(5)}R^{ij}=\dis\frac{1}{2}b^{[i}\wedge h^{j]}\inn(b^m\wedge F_m)\,,
   \qquad  &{}^{(1)}R^{ij}=R^{ij}-\sum_{a=2}^6{}^{(a)}R^{ij}\, .\lab{B.2a}
\ea
\ee
where
\bea
&&F^i:=h_m\inn R^{mi}=\ric^i\, , \qquad F:=h_i\inn F^i=R\, ,    \nn\\
&&X^i:={}^*(R^{ik}\wedge b_k)\, ,\qquad X:=h_i\inn X^i\,.       \lab{B.2b}
\eea
and
\bea
&&\Phi_i:=F_i-\frac{1}{4}b_iF-\frac{1}{2}h_i\inn(b^m\wedge F_m)\,,\nn\\
&&\Psi_i:=X_i-\frac{1}{4}b_i X-\frac{1}{2}h_i\inn(b^m\wedge X_m)\, .
\eea
\esubeq

The above formulas differ from those in Refs. \cite{x4,x22} in two minor
details: the definitions of $F^i$ and $X^i$ are taken with an additional
minus sign, but at the same time, the overall signs of all the irreducible
curvature parts are also changed, leaving their final content unchanged.

\section{Calculating the PGT field equations}\label{appC}
\setcounter{equation}{0}

The gravitational dynamics of PGT is determined by a Lagrangian
$L_G=L_G(b^i,T^i,R^{ij})$ (4-form), which is assumed to be at most
quadratic in the field strengths (quadratic PGT) and parity invariant
\cite{x23}. The form of $L_G$ can be conveniently represented as
\be
L_G=-\hd(a_0R+2\L) +\frac{1}{2}T^iH_i+\frac{1}{4}R^{ij}H'_{ij}\,,\lab{C.1}
\ee
where $H_i:=\pd L_G/\pd T^i$ (the covariant momentum) and $H'_{ij}$ define
the quadratic terms in $L_G$:
\bsubeq\lab{C.2}
\be
H_i=2\sum_{n=1}^3\hd(a_n{}^{(n)}T_i)\, ,\qquad
H'_{ij}:=2\sum_{n=1}^6\hd(b_n{}^{(n)}R_{ij})\, .
\ee
Varying $L_G$ with respect to $b^i$ and $\om^{ij}$ yields the PGT field
equations in vacuum. After introducing the complete covariant momentum
$H_{ij}:=\pd L_G/\pd R^{ij}$ by
\be
H_{ij}=-2a_0\hd(b^ib^j)+H'_{ij}\, ,
\ee
\esubeq
these equations can be written in a compact form as \cite{x4,x22}
\bea
(1ST)&& \nab H_i+E_i=0\, ,                                      \nn\\
(2ND)&& \nab H_{ij}+E_{ij}=0\, ,                                \lab{C.3}
\eea
where $E_i$ and $E_{ij}$ are the gravitational energy-momentum and spin
currents:
\bea
&&E_i:=h_i\inn L_G-(h_i\inn T^m)H_m
                 -\frac{1}{2}(h_i\inn R^{mn})H_{mn}\, ,         \nn\\
&&E_{ij}:=-(b_iH_j-b_j H_i)\, .
\eea

The above procedure is used in subsection \ref{sub32} to find explicit
form of the PGT field equations for the \ppl\ waves with torsion, with the
result displayed in Eqs. \eq{3.6}, \eq{3.7} and \eq{3.8}. To simplify
calculation of the term $\nab\hd{}^{(1)}R_{ij}$ in $\nab H_{ij}$, we used
the identity
\be
\frac{1}{2}\nab\hd R_{ij}=\nab\hd{}^{(2)}R_{ij}+\nab\hd{}^{(4)}R_{ij}\,,
\ee
that follows from the Bianchi identity $\nab R^{ij}=0$ and the double
duality properties of the irreducible parts of the curvature.


\begin{thebibliography}{99}

\bibitem{x1} H. Weyl, Electron and gravitation, I. (in German),
    Zeitschrift f\"ur Physik, {\bf 56} (1929) 330--352, translated in: L.
    O'Raifeartaigh, \emph{The Dawning of Gauge Theory} (Princeton
    University Press, Princeton, 1997), pp. 121--144.

\bibitem{x2} T. W. B. Kibble, Lorentz invariance and the gravitational
    field, J. Math. Phys. {\bf 2}, 212--221 (1961);

    D. W. Sciama, On the analogy between charge and spin in general
    relativity, in: Recent Developments in General Relativity, Festschrift
    for Infeld (Pergamon Press, Oxford; PWN, Warsaw, 1962) pp. 415--439.

\bibitem{x3} M. Blagojevi\'c, \emph{Gravitation and Gauge Symmetries}
    (IoP Publishing, Bristol, 2002);

    T. Ort\'yn, \emph{Gravity and Strings} (Cambridge University Press,
    Cambridge, 2004).

\bibitem{x4} Yu. N. Obukhov, Poincar\'e gauge gravity: Selected topics,
    Int. J. Geom. Meth. Mod. Phys. {\bf 3}, 95--138 (2006).

\bibitem{x5} M. Blagojevi\'c and F. W. Hehl (eds.), \emph{Gauge Theories
    of Gravitation, A Reader with Commentaries} (Imperial College Press,
    London, 2013).

\bibitem{x6} J. Ehlers and W. Kundt, Exact solutions of the gravitational
    field equations, in: \emph{Gravitation: an Introduction to Current
    Research}, ed. L. Witten (Willey, New York, 1962) pp. 49--101.

\bibitem{x7} V. Zakharov, \emph{Gravitational Waves in Einstein's Theory}
    (Halsted Press, New York, 1973).

\bibitem{x8} H. Stephani, D. Kramer, M. MacCallum, C. Hoenselaers, and E.
    Herlt, \emph{Exact Solu\-tions to Einstein's Field Equations}, 2nd ed.
    (Cambridge University Press, Cambridge, 2003).

\bibitem{x9} J. B. Griffiths and J. Podolsk\'y, \emph{Exact Space-Times in
    Einstein's General Relativity}, (Cambridge University Press,
    Cambridge, 2009).

\bibitem{x10} J. Bi\v c\'ak and J. Podolsk\'y, Gravitational waves in
    vacuum spacetimes with cosmological constant: I. Classification and
    geometrical properties of nontwisting type N solutions, J. Math. Phys.
    {\bf 40}, 4495--4505 (1999).

\bibitem{x11} J. B. Griffiths, P. Docherty and J. Podolsk\'y, Generalized
    Kundt waves and their physical interpretation, Class. Quant. Grav.
    {\bf 21}, 207--222 (2004).

\bibitem{x12} I. Osv\'ath, I. Robinson and K. R\'ozga, Plane-fronted
    gravitational and electromagnetic waves in spaces with cosmological
    constant, J. Math. Phys. {\bf 26}, 1755-1761 (1985).

\bibitem{x13}  M. Blagojevi\'c and B. Cvetkovi\'c, Siklos waves in
    Poincar\'e gauge theory, Phys. Rev. {\bf D} 93 (2016) 044018 (9
    pages); see also: Siklos waves with torsion in 3D, JHEP {\bf 11}, 141
    (2014) [16 pages].

\bibitem{x14} W. Adamowicz, Plane waves in gauge theories of gravitation,
    Gen. Rel. Grav. 12, 677--691 (1980);

    M.-K. Chen, D.-C. Chern, R.-R. Hsu, and W. B. Yeung, Plane-fronted
    torsion waves in a gravitational gauge theory with a quadratic
    Lagrangian, Phys. Rev. D {\bf 28}, 2094--2095 (1983);

    R. Sippel and H. Goenner, Symmetry classes of pp-waves, Gen. Rel.
    Grav. {\bf 18}, 1229--1243 (1986);

    P. Singh, On null tratorial torsion in vacuum quadratic Poincar\'e
    gauge theory, Class. Quant. Grav. {\bf 7}, 2125--2130 (1990);

    P. Singh and J. B. Griffiths, A new class of exact solutions of the
    vacuum quadratic Poincare gauge field theory, Gen. Rel. Grav. {\bf
    22}, 947--956 (1990);

    V. V. Zhytnikov, Wavelike exact solutions of $R + R^2 + Q^2$ gravity,
    J. Math. Phys. {\bf 35}, 6001--6017 (1994);

    O. V. Babourova, B. N. Frolov and E. A. Klimova, Plane torsion waves
    in quadratic gravitational theories, Class. Quant. Grav. {\bf 16},
    1149--1162 (1999);

    A. D. King and D. Vassiliev, Torsion waves in metric-affine field
    theory, Class. Quantum Grav. {\bf 18}, 2317--2329 (2001);

    V. Pasic and D. Vassiliev, PP-waves with torsion and metric-affine
    gravity, Class. Quant. Grav. {\bf 22}, 3961--3976 (2005).

    V. Pasic and E. Barakovic, PP-waves with torsion: a metric-affine
    model for the massless neutrino, Gen. Rel. Grav. {\bf 46}, 1787 (2014)
    [27 pages].

\bibitem{x15} Yu. N. Obukhov, Generalized plane-fronted gravitational
    waves in any dimension, Phys. Rev. D {\bf 69}, 024013 (2004).

\bibitem{x16} The field equations \eq{3.6} for the \ppl\ waves with
    torsion are checked using the Excalc package of the computer algebra
    system Reduce; after being transformed to the form \eq{3.8}, they are
    solved with the help of Wolfram Mathematica.

\bibitem{x17} K. Hayashi and T. Shirafuji, Gravity from Poincar\'e gauge
    theory of fundamental interactions. I, General formulation, Prog.
    Theor. Phys. {\bf 64}, 866--882 (1980); IV, Mass and energy of particle
    spectrum,  Prog. Theor. Phys. {\bf 64}, 2222--2241 (1980).

\bibitem{x18} E. Sezgin and P. van Nieuwenhuizen, New ghost-free gravity
    Lagrangians with propagating torsion, Phys. Rev. {\bf 21}, 3269--3280
    (1980);

    E. Sezgin, Class of ghost-free gravity Lagrangians with massive or
    massless propagating modes, Phys. Rev. {\bf 24}, 1677--1680 (1981).

\bibitem{x19}  G. E. Andrews, R. Askey, and R. Roy, \emph{Special
    functions} (Cambridge University Press, Cambridge, 1999);

    Z. X. Wang and D. R. Guo, \emph{Special Functions} (World Scientific,
    Singapore, 1989).

\bibitem{x20} R. M. Wald, \emph{General Relativity} (The University of
    Chicago Press, Chicago, 1984).

\bibitem{x21} S. W. Hawking and  G. F. R. Ellis, \emph{The large Scale
    Structure of Spacetime} (Cambridge University Press, 1973).

\bibitem{x22} F. W. Hehl, J. D. McCrea, E. W. Mielke, and Y. Neeman,
    Metric-affine gauge theory of gravity: Field equations, Noether
    identities, world spinors, and breaking of dilation invariance, Phys.
    Rept. {\bf 258}, 1--171 (1995).

\bibitem{x23} Y. N. Obukhov, Gravitational waves in Poincar]\'e gauge
    gravity theory, unpublished work (February 2017). The author studies
    the plane-fronted gravitational waves using the most general quadratic
    PGT Lagrangian with both parity even and parity odd terms, but
    assuming $\L=0$.


\end{thebibliography}
\end{document}